\documentclass[12pt]{article}
\usepackage{graphicx}

\newcommand{\beq}{\begin{equation}}
\newcommand{\eeq}{\end{equation}}
\newcommand{\bea}{\begin{eqnarray}}
\newcommand{\ena}{\end{eqnarray}}
\newcommand{\DFLUX}
{\mbox{$\rm cm^{-2} \; s^{-1} \; sr^{-1} \; GeV^{-1}$}} 

\newcommand{\pbar}{\ensuremath{\bar{\rm p}}}
\newcommand{\nbar}{\ensuremath{\bar{\rm n}}}
\newcommand{\dbar}{\ensuremath{\bar{\rm D}}}
\newcommand{\tbar}{\ensuremath{\bar{\rm T}}}
\newcommand{\hebar}{\ensuremath{^{3}{\bar{\rm He}}}}
\newcommand{\kchi}{\ensuremath{\vec{k_{\chi}}}}
\newcommand{\kpbar}{\ensuremath{\vec{k_{\pbar}}}}
\newcommand{\knbar}{\ensuremath{\vec{k_{\nbar}}}}
\newcommand{\kdbar}{\ensuremath{\vec{k_{\dbar}}}}
\newcommand{\ktbar}{\ensuremath{\vec{k_{\tbar}}}}
\newcommand{\kdif}{\ensuremath{\vec{\Delta}}}
\newcommand{\STOT}{\ensuremath{\sigma^{\rm tot}_{\rm p-p}}}
\newcommand{\mpt}{\ensuremath{m_{\rm p}}}

\newcommand{\md}{\ensuremath{m_{\dbar}}}
\newcommand{\mt}{\ensuremath{m_{\tbar}}}
\newcommand{\pc}{\ensuremath{P_{\rm coal}}}

\newcommand{\np}{\ensuremath{n_{\rm p}}}
\newcommand{\nC}{\ensuremath{n_{\chi}}}
\newcommand{\Ep}{\ensuremath{E_{\rm p}}}
\newcommand{\EC}{\ensuremath{E_{\chi}}}
\newcommand{\Npi}{\ensuremath{N_{{\rm p} \, i}}}
\newcommand{\Api}{\ensuremath{A_{{\rm p} \, i}}}
\newcommand{\BCi}{\ensuremath{B_{\chi \, i}}}

\begin{document}

\begin{titlepage}
\rightline{ENSLAPP-A-643/97}

\vskip 1.5cm
\begin{center}
{\Large \bf The Production of Anti-Matter in our Galaxy}
\vskip 0.5cm
Pascal Chardonnet$^{\rm a,b}$,
Jean Orloff$^{\rm a}$ and
Pierre Salati$^{\rm a,b}$
\end{center}

\vskip 0.5cm
\begin{flushleft}
{\it
a) Laboratoire de Physique Th{\'e}orique ENSLAPP, BP110, F-74941
Annecy-le-Vieux Cedex, France.\\
b) Universit{\'e} de Savoie, BP1104 73011 Chamb{\'e}ry Cedex,
France.}
\end{flushleft}

\vskip 2.cm
\begin{abstract}
The discovery of a single anti-helium nucleus in the cosmic ray flux
would definitely point toward the existence of stars and even of entire
galaxies made of anti-matter. The presence of  anti-nuclei in cosmic
rays has actually profound implications on the fundamental question
of the baryon asymmetry of the universe. It is therefore crucial to
determine the amount of anti-matter which our own galaxy already
produces through the spallation of high-energy protons on the interstellar
gas of the galactic disk. We have used here a coalescence model to assess
the amount of anti-deuterium and anti-helium $\hebar$ present in
cosmic rays together with anti-protons. The propagation of cosmic rays
in the galaxy is described through a two-zone diffusion model which correctly
describes the observed abundances. We find that the $\dbar / {\rm p}$ ratio
exceeds $10^{-9}$ above a momentum per anti-nucleon of $\sim$ 4 GeV/c.
Would the universe be purely made of matter, the AMS collaboration
should be able to detect a few anti-deuterons during the space station stage
of the experiment. However, the $\hebar / {\rm p}$ abundance does not exceed
$\sim 4 \times 10^{-13}$. Heavier anti-nuclei are even further suppressed.
\end{abstract}
\end{titlepage}

\section{Introduction.}
\label{sec:introduction}

The amount of anti-matter in cosmic rays is about to be measured with
unequalled accuracy by the space shuttle borne spectrometer of the
AMS collaboration \cite{AMS}. One of the most exciting goals of the
experiment is the possible detection of anti-nuclei in the cosmic radiation.
It is generally believed that the observation of a single anti-helium or anti-carbon
would undoubtedly signal the presence of stars made of anti-matter. Such a
discovery would be of paramount importance as regards the existence of a baryon
symmetry in the universe and has therefore strong cosmological implications.

That is why it is crucial to ascertain that cosmic rays do not already contain
detectable traces of anti-nuclei which could have been directly manufactured
in our galaxy. We know for instance that a fraction
$\pbar / {\rm p} \sim 2 \times 10^{-4}$ of anti-protons is produced by the
spallation of cosmic ray protons on the interstellar gas of the galactic disk.
In this letter, we compute the abundance of anti-deuterium $\dbar$ and
anti-helium $\hebar$ produced through the interaction of high-energy
protons with the interstellar hydrogen.

That calculation requires two ingredients. First, we need to evaluate the
production cross section of anti-nuclei during the interaction of a
high-energy proton with a proton at rest. Anti-deuterons are formed
by the fusion of an anti-proton and anti-neutron pair. We have used here
a factorization scheme together with a coalescence model \cite{Butler63}
which we discuss in section~\ref{sec:coalescence}. Our estimates of the
probability for anti-deuterium production are in fairly good agreement
with the accelerator data collected at
Serpukhov and at the ISR \cite{Albrow75,Gibson78,Abramov87}.
The anti-helium $\hebar$ is predominantly formed through the production
of anti-tritium $\tbar$ which subsequently decays into $\hebar$.
Then, in section~\ref{sec:diffusion}, we recall the salient features
of the two-zone diffusion model that takes care of the propagation of anti-nuclei
in the galaxy from the production regions to the earth. Finally, our estimates
of the cosmic ray abundances $\dbar / {\rm p}$ and $\hebar / {\rm p}$ are
discussed in section~\ref{sec:conclusion}. We find that the $\dbar / {\rm p}$ ratio
exceeds $10^{-9}$ above a momentum per anti-nucleon of $\sim$ 4 GeV/c.
Would the universe be purely made of matter, the AMS collaboration
should still be able to detect a few anti-deuterons during the space station stage
of the experiment. However, the $\hebar / {\rm p}$ abundance does not exceed
$\sim 4 \times 10^{-13}$. Heavier anti-nuclei are even further suppressed.

\section{Factorization and coalescence.}
\label{sec:coalescence}

Our goal at this point is to establish the Lorentz-invariant cross section for the
production of the anti-nucleus $\chi$ during the interaction between two
protons. Unless stated otherwise, the subsequent reasoning will be performed
in the center of mass frame of the proton-proton collision with total available
energy $\sqrt{s}$ and total cross section $\STOT$. The number
$d{\cal N}_{\chi}$ of species $\chi$ created during a single interaction and whose
momenta are $\kchi$ is related to the corresponding differential probability
of production by
\beq
d{\cal N}_{\chi} \; = \; {\cal F}_{\chi} (\sqrt{s} , \kchi) \, d^{3} \kchi \;\; .
\eeq
This probability ${\cal F}_{\chi}$ may be expressed in turn as a function
of the Lorentz-invariant differential cross section
\beq
E_{\chi} \, \frac{d^{3} \sigma_{\chi}}{d^{3} \kchi} (\sqrt{s} , \kchi) \; = \;
E_{\chi} \, {\cal F}_{\chi} (\sqrt{s} , \kchi) \, \STOT \;\; ,
\eeq
which we want to evaluate. The anti-nucleus energy $E_{\chi}$ cannot exceed
the upper bound
\beq
E_{\chi}^{\rm max} \; = \;
\frac{s - M_{X}^{2} + m_{\chi}^{2}}{2 \sqrt{s}} \;\; ,
\eeq
where $M_{X}$ denotes the minimum mass of the nucleon system (X) that
is created with the anti-nucleus. As a result of the conservation of the baryon
number, each anti-nucleon is actually forced to be produced together with
an additional nucleon. When an anti-proton is created, for instance, three
protons remain in the final state and $M_{X} = 3 \mpt$. As regards the
anti-deuterium production, $M_{X} \simeq 4 \mpt$ while it is $5 \mpt$ for
anti-tritium.

The invariant cross section for the production of anti-protons is experimentally
well known. It is fairly well fitted by the Tan and Ng's parameterization
\cite{TanNg82} which we have used throughout this analysis. Assuming that
the invariance of isospin holds, the anti-neutron production cross section is
equal to its anti-proton counterpart. The calculation of the probability for the
formation of an anti-nucleus can now be performed in two steps. We first need
to estimate the probability for the creation of a group of anti-nucleons. Then,
those anti-nucleons fuse together to yield an anti-nucleus.
Let us concentrate on the case of anti-deuterons which requires the formation of
both an anti-proton and an anti-neutron. We have assumed that factorization
holds at this stage. This means that the production of two anti-nucleons is
proportional to the square of the production of one of them, a hypothesis
which is reasonably well established at high energies.  However, at lower energies,
factorization has to break down, if only to respect the kinematic constraints that
thresholds are different. We propose to take this into account by assuming in addition
that the center of mass energy available for the production of the second anti-nucleon
is reduced by twice the energy carried away by the first anti-nucleon
\beq
{\cal F}_{\pbar , \nbar} (\sqrt{s} , \kpbar , \knbar) \; = \; \frac{1}{2} \,
{\cal F}_{\pbar} (\sqrt{s} , \kpbar) \,
{\cal F}_{\nbar} (\sqrt{s} - 2 E_{\pbar} , \knbar) \; + \;
\left( \kpbar \leftrightarrow \knbar \right) \;\; .
\label{factorization}
\eeq
This probability has the merit of vanishing at the physical threshold.
However, it introduces the arbitrary Ansatz that the first pair is produced back
to back in the center of mass frame of the proton-proton collision. To test how
crucial this assumption really is in the final results, we also give plots for a
second possible Ansatz, where the nucleon associated to the first anti-nucleon
is produced at rest. This simply amounts to replace
$\sqrt{s} - 2 E_{\pbar}$ by $\sqrt{s} - \mpt - E_{\pbar}$ in the previous formula.
As may be already guessed from figure~{\^E}\ref{fig:fig2}, both Ansaetze give similar
results. In the plateau regime where the anti-nuclei abundances are fairly
independent of the energy, the relative difference between the two
assumptions does not exceed 4\%. The probability for the production of
the $\pbar - \nbar$ pair is related to the corresponding cross section by
\beq
\left( \frac{1}{\STOT} \right) \,
\frac{d^{6} \sigma_{\pbar , \nbar}}{d^{3} \kpbar \, d^{3} \knbar}
(\sqrt{s} , \kpbar , \knbar) \; = \;
{\cal F}_{\pbar , \nbar} (\sqrt{s} , \kpbar , \knbar) \;\; .
\eeq
Although ${\cal F}$'s are not Lorentz-invariant, this relation allows to check
the Lorentz invariance of the factorization Ansatz (\ref{factorization}).

Once the anti-proton and the anti-neutron are formed, they combine
together to give an anti-deuteron with probability
\beq
{\cal F}_{\dbar} (\sqrt{s} , \kdbar) \, d^{3} \kdbar \; = \;
{\displaystyle \int} \, d^{3} \kpbar \, d^{3} \knbar \;
{\cal C}(\kpbar , \knbar) \;
{\cal F}_{\pbar , \nbar} \left( \sqrt{s} , \kpbar , \knbar \right) \;\; .
\label{coalescence_1}
\eeq
The summation is performed on those anti-nucleon configurations for which
\beq
\kpbar + \knbar \; = \; \kdbar \;\; .
\eeq
The coalescence function ${\cal C}(\kpbar , \knbar)$ describes the probability
for a $\pbar - \nbar$ pair to yield by fusion an anti-deuteron. That function
depends actually on the difference
$\kpbar - \knbar = 2 \kdif$ between the anti-nucleon momenta so that
relation (\ref{coalescence_1}) may be expressed as
\beq
{\cal F}_{\dbar} (\sqrt{s} , \kdbar) \; = \;
{\displaystyle \int} \, d^{3} \kdif \; {\cal C}(\kdif) \;
{\cal F}_{\pbar , \nbar} \left( \sqrt{s} ,
\kpbar \, = \, \frac{\kdbar}{2} + \kdif ,
\knbar \, = \, \frac{\kdbar}{2} - \kdif \right) \;\; .
\label{coalescence_2}
\eeq
Notice that the formation of the anti-nucleon pair is implicitly assumed
to be independent of the later coalescence process where the anti-nucleons
melt together to form the anti-deuteron \cite{Butler63}. This is well justified
by the large difference between the binding and pair creation energies. As a
matter of fact, an energy of $\sim 3.7$ GeV is required to form an anti-deuteron
whereas the binding energy of the latter is $B \sim 2.2$ MeV. The coalescence
function is therefore strongly peaked around $\kdif = \vec{0}$ and expression
(\ref{coalescence_2}) simplifies into
\beq
{\cal F}_{\dbar} (\sqrt{s} , \kdbar) \; \simeq \; \left\{
{\displaystyle \int} \, d^{3} \kdif \; {\cal C}(\kdif) \right\} \;
{\cal F}_{\pbar , \nbar} \left( \sqrt{s} ,
\kpbar \, = \, \frac{\kdbar}{2} , \knbar \, = \, \frac{\kdbar}{2}
\right) \;\; ,
\eeq
where the probability for the formation of the $\pbar - \nbar$ pair
has been factored out. From the term in brackets, we build a
Lorentz-invariant quantity which we evaluate this time in
the rest frame of the anti-deuteron
\beq
{\displaystyle \int} \, \frac{E_{\dbar}}{E_{\pbar} \, E_{\nbar}} \,
d^{3} \kdif \, {\cal C}(\kdif) \; \simeq \;
\left( \frac{\md}{m_{\pbar} \, m_{\nbar}} \right) \,
\left(\frac{4}{3} \pi \pc^{3} \right) \;\; .
\eeq 
In that frame, the anti-nucleons merge together if the momentum of the
corresponding two-body reduced system is less than some critical value
$\pc$. That coalescence momentum is the only free parameter of our
factorization and coalescence model. The invariant cross section for
anti-deuterium production in proton-proton collisions may finally be
expressed as
\bea
\label{anti_deuteron}
\lefteqn{
E_{\dbar} \, \frac{d^3 \sigma_{\dbar}}{d^{3} \kdbar} \; = \;
\left( \frac{\md}{m_{\pbar} \, m_{\nbar}} \right) \,
\left(\frac{4}{3} \pi \pc^{3} \right) \times
} \\
& & \frac{1}{2 \STOT} \, \left\{
E_{\pbar} \frac{d^{3} \sigma_{\pbar}}{d^{3} \kpbar}
\left( \sqrt{s} , \kpbar \right) \,
E_{\nbar} \frac{d^{3} \sigma_{\nbar}}{d^{3} \knbar}
\left( \sqrt{s} - 2 E_{\pbar} , \knbar \right)
\; + \;  \left( \kpbar \leftrightarrow \knbar \right)
\right\} \;\; ,  \nonumber
\ena
with $\kpbar = \knbar = \kdbar / 2$. The dependence on $\pc$ comes
only from the multiplicative $\pc^{3}$ prefactor.

\begin{figure}[htbp]
\begin{center}
\leavevmode
\includegraphics[width=\textwidth]{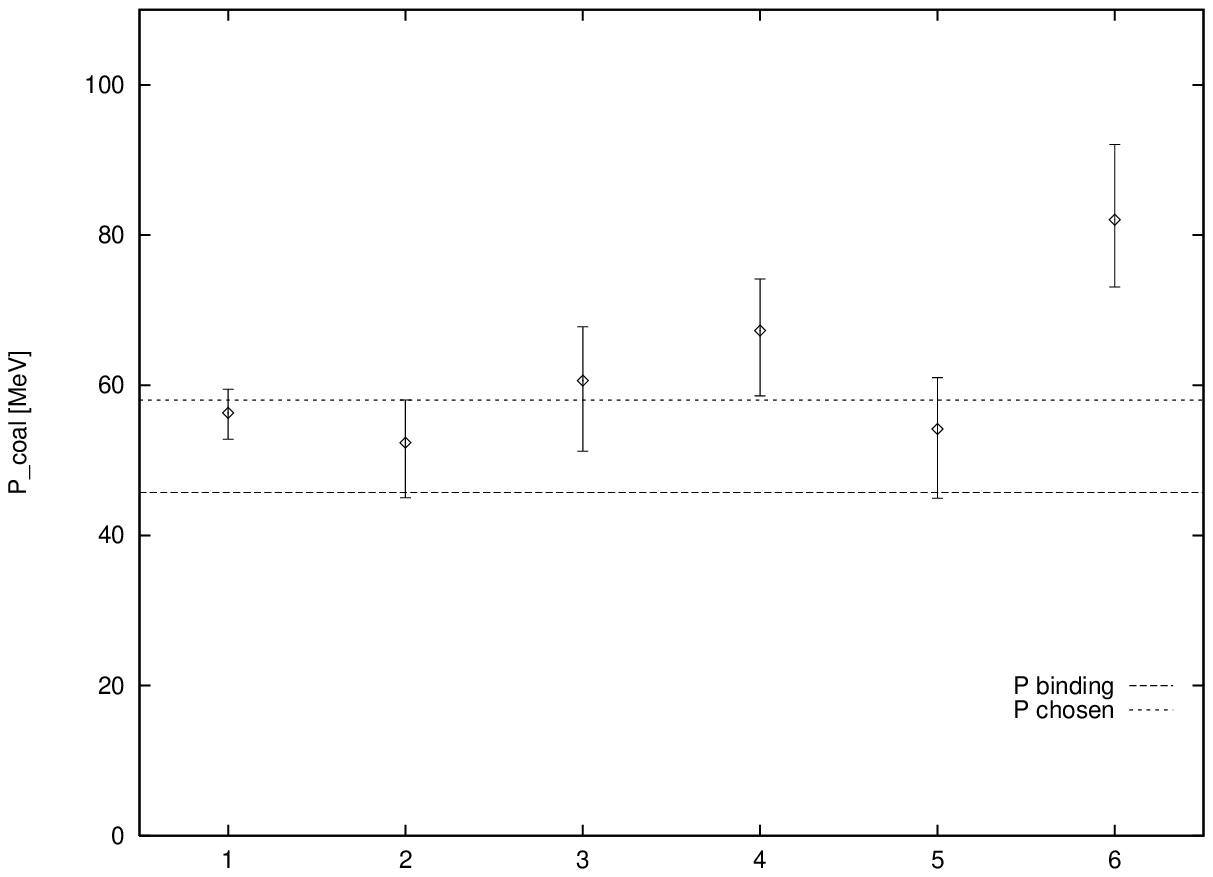}
\caption{This figure displays various experimental constraints on the
coalescence momentum \pc, the only free parameter of the model discussed
in section~(\ref{sec:coalescence}). Points 1 and 2 are from Serpukhov with
$\protect\sqrt{s}$= 11.5 GeV while all the other data have been collected at
the ISR at $\protect\sqrt{s}$ =  53 GeV. A coalescence momentum {\pc} of
order 60 MeV provides a reasonable fit of all the points but the last one.}
\label{fig:fig1}
\end{center}
\end{figure}

Theoretical values for $\pc$ range from $\sqrt{\mpt \, B} \sim 46$ MeV, naively
derived from the anti-deuteron binding energy, up to 180 Mev as would follow from
a Hulthen parametrization of the deuterium wave function \cite{Braun82}.
We therefore expect $\pc$ to lie somewhere in the range between 50 and 200
MeV. Inside this range, since factorization might also involve an unknown
coefficient that could be reabsorbed into $\pc$, we will rather follow a
phenomenological approach. We will determine $\pc$ directly from the known
experimental results. These have been summarized in terms of $\pc$ on
figure~\ref{fig:fig1}. Points 1 and 2 are from an experiment carried out at
Serpukhov \cite{Abramov87}, for a center of mass (CM) energy 
$\sqrt{s} = 11.5$ GeV, at practically vanishing longitudinal CM momentum
$p_{L}$ and at large transverse momentum with $p_{\perp} = 1.15$ and 1.5 GeV
respectively.
Points 3, 4 and 5 have been obtained at the ISR \cite{Albrow75}, with
$\sqrt{s} = 53$ GeV, small $p_{\perp}$ (0.16, 0.21 and 0.30 GeV) and
larger $p_{L}$ (4.8, 5.7 and 8 GeV).
Finally point 6 summarizes about 8 ISR data points \cite{Gibson78}, at vanishing
$p_{L}$ and with $p_{\perp}$ ranging from 0.2 to 1 GeV. Given the crudeness of our
model and the wide range of kinematic regimes explored experimentally, it is quite
comforting that all these data are compatible within 2 standard deviations with our
predictions from our single parameter model.
Notice in that respect that most of the error bars simply ignore systematic
errors. The first five data points are all fitted with a coalescence momentum of order
60 MeV. Because the Serpukhov data correspond to a low center of mass energy, a
kinematic regime where anti-deuteron production is astrophysically predominant,
we have decided to fix the value of $\pc$ at 58 MeV. Increasing this value to 75 MeV
would simply double the production of anti-deuterium.

The anti-helium $\hebar$ nucleus may be formed directly by the fusion of
two anti-protons and an anti-neutron. A second possibility is the synthesis
of anti-tritium $\tbar$ which subsequently decays into $\hebar$ with a
half-lifetime of $\sim 12.3$ years. This second mechanism is dominant. It does
not suffer from the electromagnetic suppression due to the Coulomb repulsion
between the two anti-protons. We have therefore ignored here the direct $\hebar$
fusion and have concentrated instead on the production of $\tbar$. The corresponding
production cross section may be obtained from the direct generalization of the
anti-deuteron calculations. This yields
\bea
\label{anti_triton}
\lefteqn{
E_{\tbar} \, \frac{d^3 \sigma_{\tbar}}{d^{3} \ktbar} \; = \;
\left( \frac{\mt}{m_{\pbar} \, m_{\nbar}^{2}} \right) \,
\left(\frac{4}{3} \pi \pc^{3} \right)^{2} \,
\left( \frac{1}{\STOT} \right)^{2} \times
} \\
& &
\left\{
E_{\pbar} \frac{d^{3} \sigma_{\pbar}}{d^{3} \kpbar}
\left( \sqrt{s} , \vec{k_{1}} \right) \,
E_{\pbar} \frac{d^{3} \sigma_{\pbar}}{d^{3} \kpbar}
\left( \sqrt{s} - 2 E_{\pbar} , \vec{k_{2}} \right) \,
E_{\pbar} \frac{d^{3} \sigma_{\pbar}}{d^{3} \kpbar}
\left( \sqrt{s} - 4 E_{\pbar} , \vec{k_{3}} \right) \,
\right\} \nonumber \;\; .
\ena
The momenta $\vec{k_{1}}$, $\vec{k_{2}}$ and $\vec{k_{3}}$ of
the three anti-nucleons are all equal to ${\ktbar} / 3$ so that their energy
${E_{\pbar}}$ is also the same. Unfortunately, there are no data available to
calibrate the coalescence factor in this case. The best we can do is to use
the value ${\pc} = 58$ MeV extrapolated from the anti-deuterium
data. We have checked that relation~(\ref{anti_triton}) does not violate the
experimental upper bounds \cite{Albrow75}.

\section{The diffusion of cosmic rays in the galaxy.}
\label{sec:diffusion}

Cosmic rays are produced when supernovae shocks accelerate the
interstellar material of the galactic plane. These high-energy nuclei propagate
in the erratic magnetic fields of the disk where they interact on the gas
to create secondary cosmic rays. The amount of secondary light elements
such as lithium, beryllium and boron (Li-Be-B) is well explained by the spallation
of primary carbon, oxygen and nitrogen nuclei. The latter spend $\sim$ 10
million years (My) in the galactic plane where they cross a column density of
$\sim$ 15 ${\rm g \, cm^{-2}}$. Quite exciting is the measurement of the
isotopic ratio between the unstable $^{10}$Be and its stable partner $^{9}$Be.
The former nucleus has a half-lifetime of 1.6 My and plays the role of
a chronometer. A low value is observed and indicates that cosmic rays
spend actually 100 My in the galaxy before escaping in outer space. Particles
are confined 90\% of the time in extended layers above and beneath the matter
ridge where they just diffuse without interacting much with the scarce interstellar
medium. Therefore, as regards the propagation of cosmic rays, our galaxy
may be reasonably well modelled with two main regions. First, cosmic rays
are accelerated within a thin disk with radius
$0 \leq r \leq R \; = \; 20$ kpc and thickness
$ |z| \leq h \; = \; 100$ pc, where they diffuse and interact on atomic and
molecular hydrogen. This gaseous plane is sandwiched by extended domains
containing irregular magnetic fields, with same radial extension and
$|z| \leq L$ = 3 kpc. These thick layers play the role of confinement reservoirs.
As a matter of fact, the presence of magnetic fields far above the galactic plane
is now firmly established by synchrotron radiation as mentioned by Badwar
\cite{BADWAR}. On average, the intensity of the magnetic field decreases from
$\sim$ 5 $\mu$G in the plane down to 1 $\mu$G at a height of 5 kpc. Note
that magnetic fields have also been detected in other galaxies \cite{SOFUE}.
Therefore, the transport of high-energy particles crucially depends on their
diffusion across the chaotic magnetic fields of the galaxy. We will assume here
an isotropic diffusion with coefficient given by the empirical value
\beq
K \; = \; 6 \times 10^{23} \; {\rm m^{2} \, s^{-1}} \,
\left( 1 +
\frac{\cal R}{\mbox{3 GV}} \right)^{0.6} \;\; ,
\label{diffusion_coefficient}
\eeq
where ${\cal R} = p / Z$ is the particle rigidity. The diffusion coefficient $K$ is
constant at low  energies, but increases with the rigidity beyond a critical value
of 3 GV.

We follow here the analysis of Webber, Lee and Gupta \cite{WLG}. The authors
showed that the two-zone diffusion model presented above is in good agreement
with the observed primary and secondary nuclei abundances. Assuming that steady
state holds, the distribution of cosmic ray protons may be derived from the diffusion
equation
\bea
\label{p_propagation}
\lefteqn{
{\displaystyle \frac{\partial \np}{\partial t}} \; = \; 0 \; =
} \\
& & = \; K \, \left\{ \frac{\partial^{2} \np}{\partial r^{2}} +
\frac{1}{r} \frac{\partial \np}{\partial r} +
\frac{\partial^{2} \np}{\partial z^{2}} \right\}
\;  - \;
2 h \, \delta(z) \Gamma_{\rm p} \, \np \; + \; 2 h \, \delta(z) Q_{\rm p}(r) \;\; ,
\nonumber
\ena
where $\np(\Ep,r,z)$ is the density of protons with energy $\Ep$ at location 
$(r,z)$. In the right-hand side of relation (\ref{p_propagation}),
the first term describes the diffusion of particles.
The second term accounts for the destruction of protons by their interactions with
the interstellar medium of the galactic plane. That destruction rate is shown to be
very small but has not been neglected in what follows. The disk is only 200 pc thick,
hence our approximation of an infinite thinness and the term $2 h \, \delta(z)$.
The collision rate of protons with the interstellar hydrogen is given by
\beq
\Gamma_{\rm p} \; = \; \STOT \; v_{\rm p} \; n_{\rm H} \;\; ,
\label{p_collision}
\eeq
where $\STOT \sim$ 44 mbarns is the total proton-proton interaction
cross section, $v_{\rm p}$ denotes the velocity and $n_{\rm H} = 1$ cm$^{-3}$ is the
average  hydrogen density in the thin matter disk.
The last term in relation 
(\ref{p_propagation}) deals with the sources of high-energy protons. It matches the
distribution of supernovae remnants and pulsars as measured by Lyne, Manchester and
Taylor \cite{LMT}
\beq
Q_{\rm p}(\Ep , r) \; = \; Q_{\rm p}^{0}(\Ep) \, \rho^{0.6} \,
e^{- 3 {\displaystyle \rho}} \;\; ,
\label{source_radius}
\eeq
with $\rho = r / R$.
At the edge of the domain where cosmic rays are confined, the particles escape freely
and diffusion becomes inefficient. Thus the density vanishes at the boundaries of the
axi-symmetric domain where cosmic rays propagate. This provides the boundary
conditions for solving the diffusion equation (\ref{p_propagation}). Following
\cite{WLG}, the density $\np$ is expanded as a series of the Bessel functions of zeroth
order $J_{0} \left( \zeta_i \rho \right)$ where $\zeta_i$ is the ith zero of $J_0$
and where $\rho = r / R$. At fixed proton energy $\Ep$ and corresponding rigidity
${\cal R}_{\rm p}$, the density may be expressed as
\beq
\np ( r , z ) \; = \; {\displaystyle \sum_{i=1}^{\infty}} \; \Npi (z) \;
J_{0} \left( \zeta_i \rho \right) \;\; .
\label{np}
\eeq
The diffusion equation (\ref{p_propagation}) may be Bessel transformed,
yielding
\beq
K \, \left\{ \frac{d^{2} \Npi}{dz^{2}} \, - \, \frac{\zeta_i^{2}}{R^{2}} \Npi \right\}
\; - \; 2 h \, \delta(z) \Gamma_{\rm p} \, \Npi
\; + \; \delta(z) q_{i} \; = \; 0 \;\; .
\label{p_propagation_i}
\eeq
The coefficients $q_{i} = 2 h \, Q_{i}$ are the Bessel transforms of the source
distribution $Q_{\rm p}(r)$ in the galactic plane
\beq
q_{i} \; = \;
\left\{ \frac{Q_{\rm p}^{\rm tot}}{\pi R^{2}} \right\}
\left\{ J_1^{-2} (\zeta_i) \right\}
\left\{ {\displaystyle \int_{0}^{1}} \, Q_{\rm p}(\rho)
J_{0} \left( \zeta_i \rho \right) \rho \, d\rho \right\}
\left\{ {\displaystyle \int_{0}^{1}} \, Q_{\rm p}(\rho) \rho \, d\rho \right\}^{-1}
\;\; ,
\eeq
where $Q_{\rm p}^{\rm tot}$ denotes the total production rate of cosmic ray protons
with energy $\Ep$ in the entire galactic ridge. We have assumed here that the energy
spectrum of the sources has the same shape all over the disk. Therefore, the energy
dependence of the coefficients $q_{i}$ may be factorized out in the term
$Q_{\rm p}^{\rm tot}(\Ep)$ and does not depend on $i$. The functions $\Npi$
are derived from the requirement that they vanish at the boundary $z = L$ and from
the relation
\beq
2 K
\left. \frac{d \Npi}{dz} \right|_{\displaystyle z=0} \; = \;
2 h \Gamma_{\rm p} \Npi (0) \, - \, q_{i} \;\; .
\eeq
Integrating equation (\ref{p_propagation_i}) across the disk and remembering
that the $\Npi$'s are odd functions of the height $z$ leads to the last expression.
The cosmic ray proton density may finally be expressed as
\beq
\Npi (z) \; = \; \frac{q_{i}}{\Api} \, {\cal F}_{i}(z) \;\; .
\label{Npi}
\eeq
The vertical distribution ${\cal F}_{i}(z)$ is given by
\beq
{\cal F}_{i}(z) \; = \; {\displaystyle \frac
{\sinh \left\{ {\displaystyle \frac{\alpha_{i}}{2}} (L - |z|) \right\}}
{\sinh \left\{ {\displaystyle \frac{\alpha_{i}}{2}} L \right\}}} \;\; ,
\eeq
where the parameters $\alpha_{i}$ are equal to $2 \, \zeta_i / R$. The
coefficients $\Api$ stand for
\beq
\Api \; = \; 2 h \Gamma_{\rm p} \; + \;
K \alpha_{i} \coth \left\{ {\displaystyle \frac{\alpha_{i}}{2}} L \right\} \;\; ,
\eeq
and depend on the proton energy $\Ep$ through the diffusion coefficient
$K ({\cal R}_{\rm p})$ of relation (\ref{diffusion_coefficient}) and, in a
lesser extent, through the collision rate
$\Gamma_{\rm p}$. These coefficients are actually dominated by the diffusion
contribution. In the case of cosmic ray protons, spallation does not play an
important role in determining their galactic distribution. It affects heavier
nuclei for which the interaction cross section on hydrogen is significantly larger.

The density $\nC$ of the anti-nucleus species $\chi$ may be readily inferred
from a similar reasoning. It satisfies the diffusion equation
\beq
K \, \Delta \nC \; - \;
2 h \, \delta(z) \Gamma_{\chi} \, \nC \; + \; 2 h \, \delta(z) Q_{\chi}(r)
\; = \; 0 \;\; .
\eeq
The source term $Q_{\chi}(r)$ obtains now from the convolution of the proton energy
spectrum $\np$ with the cross section for production of an anti-nucleus $\chi$ in a
proton-proton interaction
\beq
Q_{\chi}(\EC , r) \; = \;
{\displaystyle \int_{E_{\rm p}^{\rm min}}^{+ \infty}} \,
\Gamma_{\rm p} \, {\np}(\Ep , r , z=0) \,
\left\{ \frac{d{\cal N}_{\chi}}{d\EC} (\Ep \to \EC) \right\} \,
d\Ep \;\; .
\eeq
The energy threshold above which the cosmic ray proton may produce an
anti-nucleus $\chi$ depends on the atomic number of the latter. For anti-protons,
$E_{\rm p}^{\rm min} = 7 \, \mpt$ while it is 17 $\mpt$ for anti-deuterons
and 31 $\mpt$ for anti-tritons.
The differential energy distribution ${d{\cal N}_{\chi}}/{d\EC}$ corresponds to
the anti-nucleus $\chi$ with energy $\EC$ that is produced when a proton with
energy $\Ep$ collides on an hydrogen atom at rest. It is related to the
Lorentz-invariant production cross section $\sigma_{\chi}$ estimated in
section~\ref{sec:coalescence}
\beq
\STOT \, \frac{d{\cal N}_{\chi}}{d\EC} \; = \;
{\displaystyle \int_{0}^{\pi}} \,
2 \pi \, k_{\chi} \,
\left\{ \EC \frac{d^{3} \sigma_{\chi}}{d^{3} \kchi} \right\} \,
d(- cos \theta) \;\; ,
\eeq
where $\sigma_{\chi}$ corresponds to the process
${\rm p + p} \to \chi + {\rm X}$. The momentum $\kchi$ of the outgoing
species $\chi$ makes an angle $\theta$ with respect to the direction of the incoming
cosmic ray proton. The production term $Q_{\chi}$ may be expanded in terms
of its Bessel transforms $S_{i}$. This yields the anti-nucleus distribution
\beq
\nC (\EC , r , z) \; = \; {\displaystyle \sum_{i=1}^{\infty}}
\, \left( \frac{2 h \, S_{i}}{\BCi} \right) \,
{\cal F}_{i}(z) \,
J_{0} \left( \zeta_i \rho \right) \;\; ,
\label{chi_propagation_a}
\eeq
where
\beq
\BCi (\EC) \; = \; \; 2 h \Gamma_{\chi} \; + \;
K({\cal R}_{\chi})
\alpha_{i} \coth \left\{ {\displaystyle \frac{\alpha_{i}}{2}} L \right\} \;\; .
\eeq
Relation (\ref{chi_propagation_a}) may be simplified because the Bessel
transforms $\Npi$ of the proton distribution $\np$ have all the same energy
behavior. The latter is contained in the factor
$Q_{\rm p}^{\rm tot}(\Ep) / K(\Ep)$ as may be inferred from expression
(\ref{Npi}). An effective anti-nucleus multiplicity may therefore be defined as
\beq
{\cal N}_{\chi}^{\rm eff} (\EC) \; = \;
{\displaystyle \int_{E_{\rm p}^{\rm min}}^{+ \infty}} \;
\left\{ \frac{\Phi_{\rm p} (\Ep)}{\Phi_{\rm p} (\EC)} \right\} \;
\left\{ \frac{d{\cal N}_{\chi}}{d\EC} (\Ep \to \EC) \right\} \,
d\Ep \;\; .
\label{multiplicity_chi}
\eeq
No convolution with the energy spectrum $\np$ of the cosmic ray protons is
needed any more. The differential flux of protons with energy $\Ep$ is denoted
here by
\beq
\Phi_{\rm p} (\Ep) \; = \; \frac{1}{4 \pi} \np v_{\rm p} \;\; ,
\eeq
and is expressed in units of $\DFLUX$. In order to derive that flux,
we could have naively computed the proton density $\np$ from the Bessel
expansion (\ref{np}), provided $Q_{\rm p}^{0}(\Ep)$ is well known.
However, as explained above, cosmic ray protons
are produced with the same energy spectrum all over the galactic ridge. Their
propagation is also dominated by mere diffusion and not spallation on the
interstellar gas. Therefore, their energy spectrum does not
depend on the position in the galaxy. The energy dependence of $\np$ may
actually be factorized. As a consequence, the flux ratio that appears in the effective
multiplicity (\ref{multiplicity_chi}) is constant throughout the galaxy and may be
directly derived from measurements performed at earth. We borrowed
the proton spectrum from \cite{GAISSER}
\beq
\Phi_{\rm p} (\Ep) \; \propto \;
\frac{{\cal R}_{\rm p}^{- 2.7}}
{\sqrt{1 \, + \, \left( 1.5 \, {\rm GV} / {\cal R}_{\rm p} \right)^{2}}} \;\; .
\eeq
The distribution of the anti-nucleus species $\chi$ may finally be expressed as
\beq
\nC (\EC , r , z) \; = \; {\displaystyle \sum_{i=1}^{\infty}}
\, {\cal F}_{i}(z) \, J_{0} \left( \zeta_i \rho \right) \;
\frac{q_{i}}{\Api \, \BCi} \;
2 h \, \Gamma_{\rm p} \, {\cal N}_{\chi}^{\rm eff} \;\; .
\eeq
Note that the anti-nucleus and proton energies are both equal to $\EC$
in the last formula. The diffusion coefficient $K$ that appears in
the quantities $\Api$ and $\BCi$ is therefore different for the proton
and for the anti-nucleus because at fixed energy, the corresponding
rigidities are different. Finally, the energy dependence $Q_{\rm p}^{0}(\Ep)$
in the proton sources $q_{i}$ disappears in the ratio $\nC / \np$.

\section{Discussion and conclusions.}
\label{sec:conclusion}

\begin{figure}[htbp]
\begin{center}
\leavevmode
\includegraphics[width=\textwidth]{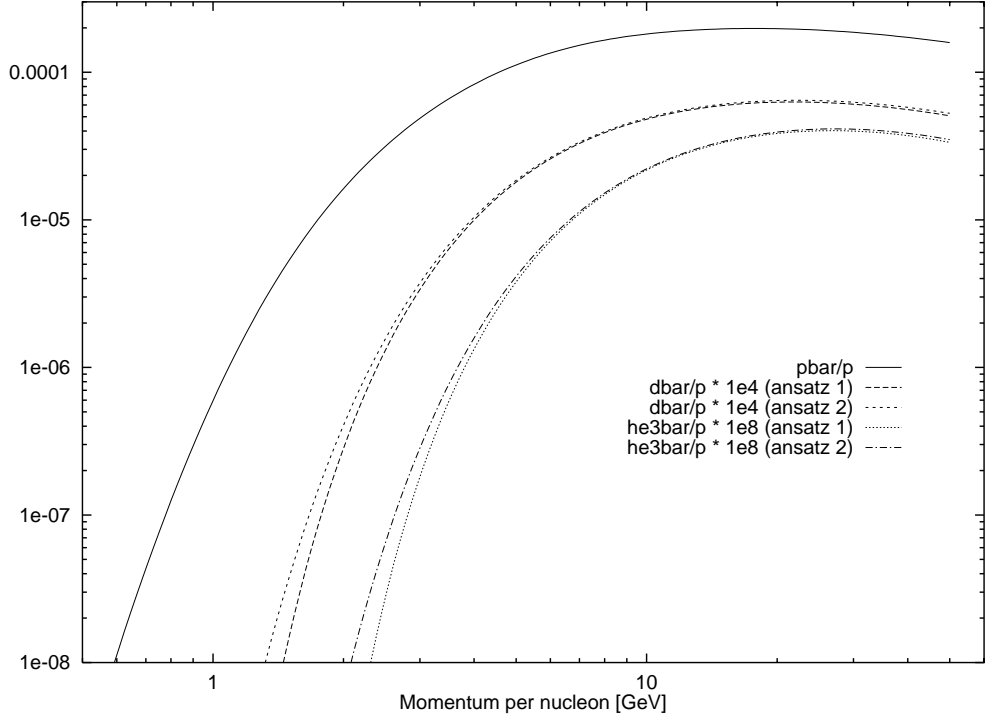}
\caption{The fluxes of cosmic ray anti-protons and of anti-deuterium and
anti-tritium nuclei, relative to the proton flux, are presented as a function
of the momentum per nucleon. To fit on the same diagram, the curves
have been scaled by a factor of $10^{4}$ for anti-deuterium and of $10^{8}$
for anti-helium $\hebar$. The doubling of curves corresponds to different
factorization schemes as explained in section~\ref{sec:introduction}.}
\label{fig:fig2}
\end{center}
\end{figure}

The flux of cosmic ray anti-protons relative to the flux of protons
is plotted (solid line) in figure~\ref{fig:fig2} as a function of the
momentum of the particles. The curve drops sharply below a few GeV.
At higher energies, it exhibits a plateau  where the ratio
$\pbar / {\rm p}$ reaches a value of $\sim 2 \times 10^{-4}$, in good
agreement with the previous calculation by Gaisser and Schaefer
\cite{GAISSER}. The $\dbar / {\rm p}$ and $\hebar / {\rm p}$
cases are also presented on the same plot. The energy behavior is
fairly the same as for the anti-protons. However, the magnitude of
the effect is significantly suppressed. To fit on the same diagram, the curves
have been scaled by a bench-mark factor of $10^{4}$ for anti-deuterium
and of $10^{8}$ for anti-helium $\hebar$. As explained in
section~\ref{sec:introduction}, the prescription that sets the energy
available to the production of any additional anti-nucleon in a
proton-proton collision is not clear.
Two options are presented here where the nucleon associated to a final
anti-nucleon is either produced back to back with its anti-partner
(lower curve) or at rest (upper curve). The effect is negligible since the
magnitude of the various plateaux is not affected.

We find that the $\dbar / {\rm p}$ ratio exceeds $10^{-9}$ above a
momentum per anti-nucleon of $\sim$ 4 GeV/c. It reaches a maximum of
$6 \times 10^{-9}$ for a momentum of $\sim$ 20 GeV/c. Note that
we have only computed the anti-deuterium yield resulting from the
spallations of cosmic ray protons on the interstellar hydrogen. Anti-deuterons
may alternatively be produced when anti-protons this time interact on the
galactic gas. We nevertheless think that this effect is negligible. On the one hand
side, the ${\rm p} \pbar$ interaction needs to be quite elastic to allow for the
survival of the incoming anti-proton. On the other hand side, if the collision
is not inelastic, the additional anti-neutron is not produced. Most of the time,
the anti-proton annihilates. The probability that a second anti-nucleon is
created with same momentum as the incoming anti-proton is therefore
extremely small. In addition, the anti-proton flux is 4 orders of magnitude
fainter as for protons.

During the space station borne stage of the experiment,
the AMS collaboration should measure a flux of anti-nucleons with a
sensitivity reaching a level of $\sim 10^{-9}$ relative to cosmic ray protons.
The corresponding energy per nucleon ranges from a GeV up to 20 GeV.
We therefore conclude that AMS should detect a few cosmic ray
anti-deuterons.

The $\hebar / {\rm p}$ abundance does not exceed $\sim 4 \times 10^{-13}$.
Even allowing for a generous error of 2 in our estimates, we conclude that
the anti-helium nuclei that are manufactured in our galaxy will not be
detected by AMS. Note that we have actually computed the interstellar flux.
The results of figure~\ref{fig:fig2} do not include solar modulation.
Since the various curves reach their maxima for a momentum per nucleon
in excess of 10 GeV, the effect of solar modulation will just be a slight
shift of a fraction of a GeV towards low energies, depending on the
specific epoch of the solar cycle. The same conclusions still hold.

As shown in section~\ref{sec:introduction}, factorization comes into play
for the production of several anti-nucleons during the same proton-proton event.
The general trend is a reduction of the yield by a factor larger than $10^{4}$
when the atomic number of the anti-nucleus is increased by one unit.
As an illustration, it is clear from figure~\ref{fig:fig2} that the $\dbar / \pbar$
ratio  is $\sim 3 \times 10^{-5}$ while $\hebar / \dbar \sim 7 \times 10^{-5}$.
The anti-nuclei with atomic number $A \geq 3$ that are produced through
cosmic ray spallations in our galaxy are not detectable by AMS.
Alternatively, if anti-matter is processed in anti-stars, the abundance of its
various components should be quite similar to the conventional stellar yields.
It would follow therefore the same relative proportions as in our own interstellar
material. In any case, the detection of a single anti-helium or anti-carbon by the
AMS collaboration would be a smoking gun for the presence of large amounts
of anti-matter in the universe and for the existence of anti-stars and of anti-galaxies.

\vskip 1.cm
\noindent {\bf Acknowledgements}
\vskip 0.5cm
\noindent We would like to express our gratitude toward G.~Girardi and
G.~Mignola for stimulating discussions. This work has been carried out under the
auspices of the Human Capital and Mobility Programme of the European Economic
Community, under contract number CHRX-CT93-0120 (DG 12 COMA). We also
acknowledge the financial support of a Collaborative Research Grant from NATO
under contract CRG 930695.


%
\newpage
\newpage
\end{document}